\documentclass{iau} 
\usepackage{graphicx}

\title[RV Tauri Distances] 
{A homogeneous distance catalogue for Galactic RV Tauri objects}

\author[S.~B~Vickers et al]   
{Shane~B~Vickers$^1$, David J Frew$^{2,3}$, Matt S Owers$^{1,4}$, Quentin A~Parker$^{1,2}$, \and Ivan S Boji\v{c}i\'{c}$^{1,2}$}

\affiliation{$^1$Department of Physics \& Astronomy, Macquarie University, North Ryde 2109, \\email:{\tt shane.vickers@mq.edu.au} \\[\affilskip]
	$^2$The University of Hong Kong, Department of Physics, Hong Kong SAR, China \\[\affilskip]
	$^3$The University of Hong Kong, Laboratory for Space Research, Hong Kong SAR, China \\[\affilskip]
	$^4$The Australian Astronomical Observatory, North Ryde, NSW, 1670, Australia}
\pubyear{2017}
\volume{323}  
\jname{Planetary Nebulae: Multi-Wavelength Probes of Stellar and Galactic Evolution}
\editors{Xiaowei Liu, Letizia Stanghellini, \& Amanda Karakas eds.}
\begin{document}

\maketitle

\begin{abstract}
 A subset of Post-AGB (PAGB) objects are the highly luminous RV Tauri variables that show similarities to Type-II Cepheids. By using a sample of known RV Tauri stars from the Magellanic Clouds we are able to determine period luminosity relationships (PLRs) in various bands that have been used to determine the luminosities of their Galactic counterparts. We have gathered all available photometry in order to generate an SED for each object and determine the total integrated flux. This total flux combined with a calculated or inferred intrinsic luminosity leads to a distance (Vickers et al. 2015). This distance catalogue has allowed us to begin to constrain the physical parameters of this poorly understood evolutionary phase and to determine links between these physical characteristics as a function of their stellar population.
\end{abstract}

\firstsection 

\section{Introduction}

The RV Tau stars are a subclass of pulsating PAGB stars named after the class prototype RV Tauri. They are luminous, variable yellow giant stars of spectral types F-K, with typical periods of 20--80 days and with alternating deep and shallow minima. The major hurdle to determining the physical characteristics of these stars is the lack of accurate distances. This is a problem that we will rectify with the release of this, the second distance catalogue that will encompass the RV Tauri stars, R Coronae Borealis stars, extreme helium stars and late thermal pulse objects. Here we will solely focus on the determination of luminosities and distances. One of our goals is to construct a volume-limited survey of PAGB stars based on our distance catalogues (Vickers et al. 2015, 2016), to compare with a similar survey for planetary nebulae (Frew et al. 2016).
\begin{figure}[]
	\vspace{-1.5cm}
	\begin{center}
		\includegraphics[width=10cm]{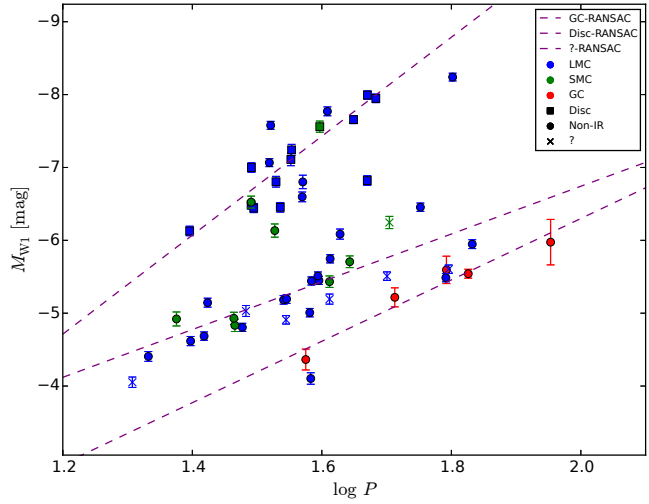} 
				\vspace{1cm}
		\caption{Period-luminosity relationship for W1 3.4 $\mu$m waveband.}
		\label{fig1}
		\vspace{-0.2cm}
	\end{center}
\end{figure}

\section{Method}
Using multi-epoch WISE photometry from the IRSA (http://irsa.ipac.caltech.edu) online database we have determined mean WISE photometry for all Galactic and extra-Galactic RV Tauri variables . The [3.4]--[4.6] WISE color is a good proxy for the presence of a dusty disk as shown in Gezer et al. (2015). We have built an SED for each RV Tauri in the Toru\'n catalogue (Szczerba et al. 2007, 2012). For this distance catalogue we have decided to use model atmospheres in a similar manner to Kamath, Wood \& Van Winckel (2015) with the exception that we have determined interstellar extinction values as described in our first PAGB distance catalogue (Vickers et al. 2015).

\section{Results}

We have constructed PLRs for each of the four WISE mid-IR wavebands using all LMC and SMC objects classified as RV Tauri stars in the OGLE-III catalogue (Soszyñski et al. 2008). Figure~1 clearly shows that the slope of the PLR for stars with a disc like SED is much steeper than those with no IR-excess. This suggests that the disc RV Tauri stars have higher intrinsic luminosities and therefore come from higher mass progenitors. The slope of the V-band PLR  shows an inflection which can be attributed to the RV Tauri stars with a large IR-excess due to a circumstellar disc (Manick et al. 2016). 

\section{Conclusion and Future Work}
We  have constructed PLRs using the LMC and SMC RV Tauri population in the mid-IR which accentuates the bending of the PLR as mentioned in Manick et al. (2016). Using these PLRs we have developed a catalogue of distances for the entire Galactic RV Tauri population. The presence of a disc increases scatter in the mid-IR PLR which must be accounted for when determining distances to Galactic RV Tauri stars.


\begin{thebibliography}{}
\bibitem[Frew \etal (2016)]{frew}
{Frew, D.J., Parker, Q.A., \& Boji\v{c}i\v{c}, I.S.} 2016
\textit{MNRAS}, 455, 1459
	
\bibitem[Gezer \etal (2015)]{gezer}
{Gezer, I., Van Winckel, H., Bozkurt, Z., De Smedt, K., Kamath, D., Hillen, M., \& Manick, R.} 2015, 
\textit{MNRAS}, 453, 133

\bibitem[Kamath \etal (2015)]{kamath_2015}
{Kamath, D., Wood, P.R., \& Van Winckel, H.} 2015, 
\textit{MNRAS}, 454, 1468

\bibitem[Manick R. \etal (2016)]{manick} 
{Manick, R., Van Winckel, H., Kamath, D., Hillen, M., \& Escorza, A.} 
\textit{A\&A} in press (arXiv:1610.00506)

\bibitem[Soszy\'nski I. \etal (2008)]{sosynski}
{Sosy\'nski, I., Udalski, A., Szyma\'nski, M.K., Kubiak, M., Pietrzy\'nski, G., Wyrzykowski, \L., Szewczyk, O., Ulaczyk, K., \& Poleski, R.} \textit{Acta Astr}, 58, 293

\bibitem[Szczerba \etal (2010)]{szczerba_2010}
{Szczerba, R., Si\'odmiak, N., Stasi\'nska, G., Borkowski, J.} 2007, 
\textit{A\&A}, 469, 799

\bibitem[Szczerba \etal (2012)]{szczerba_2012}
{Szczerba, R., Si\'odmiak, N., Stasi\'nska, G., Borkowski, J., Garc\'ia-Lario, P., Su\'arez, O., Hajduk, M., \& García-Hern\'andez, D.A.} 2012, 
in Proc. IAU Symp., 283, 506 

\bibitem[Vickers \etal (2015)]{vickers}
{Vickers, S.B., Frew, D.J., Parker, Q.A., \& Boji\v{c}i\v{c}, I.S.} 2015, 
\textit{MNRAS}, 447, 1673

\bibitem[Vickers \etal (2016)]{vickers_pacific_rim}
{Vickers, S.B., Frew, D.J., Owers, M.S., Parker, Q.A., \& Boji\v{c}i\v{c}, I.S.} 2016, 
\textit{JPhCS}, 728, 072013

\end{thebibliography}
\end{document}